\documentclass{SCIS2026}

\begin{document}
\ArticleType{RESEARCH PAPER}
\Year{2025}
\Month{January}
\Vol{68}
\No{1}
\DOI{}
\ArtNo{}
\ReceiveDate{}
\ReviseDate{}
\AcceptDate{}
\OnlineDate{}
\AuthorMark{}
\AuthorCitation{}

\title{Quantum metrology via mitigation of single-photon loss using an engineered nonlinear oscillator}{Yang T L, Ning W, Yang Z B, et al. Quantum metrology via mitigation of single-photon loss using an engineered nonlinear oscillator}

\author[1,2]{Tian-Le Yang}{}
\author[1,2]{Wen Ning}{}
\author[1,2,3]{Zhen-Biao Yang}{{zbyang@fzu.edu.cn}}
\author[1,2,3]{Shi-Biao Zheng}{{t96034@fzu.edu.cn}}


\address[1]{Fujian Key Laboratory of Quantum Information and Quantum Optics, Fuzhou University, Fuzhou 350108, China}
\address[2]{Department of Physics, Fuzhou University, Fuzhou 350108, China}
\address[3]{Hefei National Laboratory, Hefei 230088, China}

\abstract{The fragility of quantum metrological advantages under loss remains a major barrier to practical quantum sensing. For a two-photon-driven (TPD) Kerr resonator (TPD-Kerr model) subject to unavoidable single-photon loss (SPL), both the quantum Fisher information gain and squeezing level exhibit hard-to-track long-lived damped oscillations, restricting useful sensing and squeezing to extremely short time windows.
We show that adding engineered two-photon loss (ETPL)---forming a TPD-Kerr-ETPL hybrid model---significantly mitigates these oscillations and converts the decay into a smooth, monotonic drop. This extends the high-sensitivity windows by over an order of magnitude.
Moreover, we reveal a temporal hierarchy of quantum resources: the initial boost in metrological sensitivity arises from Gaussian squeezing, while sustained high-precision sensing stems from dissipatively stabilized non-Gaussian even-parity cat states.
Crucially, only in models that include ETPL---such as the TPD-Kerr-ETPL and TPD-ETPL systems---does the dynamics actively mitigate SPL's detrimental effects, transforming damped oscillation into a smooth, easily trackable trajectory and enabling a prolonged, usable metrological window.
Our approach transcends encoding-based or feedback-controlled schemes, offering a fully autonomous route to high-precision measurement without real-time feedback control. This establishes a general design principle: engineered loss, combined with appropriate driving, can actively preserve metrologically useful non-Gaussian quantum resources even in the presence of SPL---paving the way toward robust, scalable quantum sensors in superconducting circuits, optomechanics, and trapped-ion platforms.
}


\maketitle

\section{\label{sec:Introduction}Introduction}

Continuous-variable quantum systems, such as optical and mechanical oscillators, provide a powerful platform for quantum-enhanced metrology~\cite{Wolf2019,McCormick2019,Wang2019,Fadel2025}. 
A central goal in continuous-variable quantum information processing is the generation of nonclassical states of bosonic modes~\cite{RevModPhys.77.513,Andersen2010,Su2020}. 
Among these, squeezed states---Gaussian states exhibiting sub-shot-noise fluctuations in one quadrature at the expense of increased noise in the conjugate quadrature---offer a proven route to surpass the standard quantum limit in precision measurements in a variety of physical systems~\cite{PhysRevA.13.2226,PhysRevD.23.1693,Walls1983,PhysRevA.102.052611,SCHNABEL20171,Lawrie2019,PhysRevA.106.062616,Gessner2020,PhysRevLett.124.173602,PhysRevLett.124.120504,Xie2022,DiCandia2023,Zhu2023,PhysRevA.108.053327,Guo2024,Marti2024,chen2024,PhysRevLett.133.040801,PhysRevLett.132.220801,PhysRevLett.134.180801,PRXQuantum.6.020301,Alushi2025,Cai2025}.

Over the past decades, squeezed state generation has been demonstrated across diverse physical platforms, including photonic systems in both optical and microwave domains~\cite{PhysRevLett.60.764,PhysRevLett.100.033602,Safavi-Naeini2013,PhysRevLett.118.223604,PhysRevA.101.012348,PRXQuantum.2.020323,Eickbusch2022,Qiu2023,Eriksson2024,Cai2025}, as well as in mechanical and acoustic phononic systems~\cite{Wollman2015,PhysRevA.91.013834,PhysRevA.98.023807,PhysRevApplied.8.054030,Bai2019,Ma2021,Youssefi2023,Gong2023,Marti2024,PhysRevLett.133.050602,PhysRevA.111.023504}.
Parametric processes remain the most widely employed mechanism for generating squeezed light~\cite{PhysRevLett.57.2520,PhysRevLett.124.173602,PhysRevA.101.012348,PhysRevA.108.033701}. 
However, practical squeezing is often limited by intrinsic nonlinearities in the system. 
As highly squeezed states possess large excitation numbers and extended phase-space distributions---particularly in the antisqueezed quadrature---they may exceed the linear operating regime of the device, leading to non-Gaussian evolution. 
This results in ``phase-space wrapping'' of the Wigner function~\cite{Gu2025}, which degrades the degree of squeezing~\cite{PhysRevApplied.8.054030,PRXQuantum.2.020323}. 
Although such non-Gaussian states can retain high sensitivity to external perturbations~\cite{Zurek2001,PhysRevLett.107.083601,Slussarenko2017,PhysRevLett.122.040503,PhysRevLett.121.160502,Giovannetti2011,Guo2024,Marti2024}, the desired metrological resources---optimal squeezing and elevated quantum Fisher information (QFI)---are typically achieved only transiently. 
Consequently, their utility in sustained quantum sensing protocols is fundamentally constrained.

Such a limitation highlights the need for metrological frameworks that account for realistic open-system dynamics~\cite{PhysRevA.102.052611,PhysRevLett.124.120504,Xie2022,Bao2022,PhysRevLett.133.040801}. 
It has further motivated the exploration of alternative loss mechanisms capable of generating and stabilizing nonclassical quantum states over extended timescales.
The role of loss in quantum systems has undergone a profound transformation: once regarded primarily as a source of decoherence, it is now recognized as a resource that can be engineered to enable quantum state stabilization, measurement, control, and error correction~\cite{PhysRevLett.77.4728,PhysRevLett.85.1762,Harrington2022,10.21468/SciPostPhysLectNotes.72}.
A key example is engineered two-photon loss (ETPL)~\cite{Gerry1993,PhysRevA.49.2785,PhysRevLett.60.1836,Mirrahimi_2014,Leghtas2015,Puri2020,Roberts2020}, which facilitates the loss of photon pairs into a reservoir, offers a strategy to remove entropy from a memory mode and thereby stabilize a logical qubit~\cite{Mirrahimi_2014,Leghtas2015,Touzard2018,Albert_2019,PhysRevA.100.033827,Lescanne2020,PhysRevA.101.043807,Ma20211789,PRXQuantum.3.020339,Berdou2023,Gautier2023,Xu2023,Reglade2024,PhysRevX.14.021019,Marquet2024}.
The stabilization of cat qubits is further facilitated by the application of a two-photon drive to the memory mode~\cite{Mirrahimi_2014,Leghtas2015,Touzard2018,Lescanne2020,Berdou2023,Reglade2024,Marquet2024}. Beyond its role in qubit stabilization, ETPL has also enabled the extension of Wigner tomography to regimes of high photon numbers~\cite{Marquet2024} and dissipative phase transitions~\cite{Bartolo2016,Beaulieu2025}.
Although ETPL has been extensively studied for its role in quantum state engineering, there is a lack of research on utilizing its potential for enhancing and sustaining quantum squeezing and quantum sensing in two-photon-driven Kerr nonlinear oscillator systems.
Recent strategies for robust quantum metrology often rely on quantum error correction-based encoding~\cite{Layden2019,Wang2022,Zhou2024} or real-time feedback control~\cite{Fallani2022}, which require ancillary resources or continuous measurement and adjustment.
In contrast, engineered dissipation offers a fully autonomous alternative---stabilizing metrological resources without real-time intervention or additional hardware overhead.

In this work, we consider a two-photon-driven (TPD) Kerr nonlinear oscillator subject to unavoidable single-photon loss (SPL)---a common noise channel in quantum hardware---with further introduced engineered two-photon loss (ETPL). This setup allows us to investigate how metrological performance depends on the interplay between Kerr nonlinearity and ETPL. Specifically, we compare three variants of the same physical setup: (i) the TPD-Kerr model without considering ETPL, where quantum resources are generated by TPD and Kerr nonlinearity~\cite{Zurek2001,PhysRevLett.107.083601,Slussarenko2017,PhysRevLett.122.040503,PhysRevLett.121.160502,Giovannetti2011,Guo2024,Marti2024}; (ii) the TPD-ETPL model without considering Kerr nonlinearity, which relies on TPD and ETPL to stabilize states; and (iii) the hybrid TPD-Kerr-ETPL model, combining both (i) and (ii) mechanisms under identical driving condition.
We find that in the TPD-Kerr model, SPL induces long-lived oscillations in sensing performance, restricting useful operation to extremely short time windows~\cite{Zurek2001,PhysRevLett.107.083601,Slussarenko2017,PhysRevLett.122.040503,PhysRevLett.121.160502,Giovannetti2011,Guo2024,Marti2024}. By contrast, adding even weak ETPL mitigates these oscillations, yielding a smooth decay and extending the high-sensitivity window by over an order of magnitude. When ETPL dominates the Kerr nonlinearity, the hybrid system's behavior converges to that of the TPD-ETPL model---despite both TPD-Kerr and TPD-ETPL schemes being capable of producing cat states of comparable size in idealized settings, they exhibit dramatically different dynamics once realistic SPL is present.
Moreover, we uncover a temporal hierarchy of quantum resources: early enhancement of metrological sensitivity arises from Gaussian squeezing, while sustained advantage stems from dissipatively stabilized non-Gaussian cat states. Crucially, our scheme operates autonomously---without encoding~\cite{Layden2019,Wang2022,Zhou2024} or real-time feedback control~\cite{Fallani2022}---establishing a general design paradigm: ETPL, combined with coherent driving, can actively preserve metrologically useful quantum resources against dominant noise, paving the way toward robust quantum metrology in superconducting circuits, optomechanics, and trapped-ion systems.

\section{\label{sec:Model}Theoretical model}

We consider the nonlinear Kerr-resonator model subjected to a TPD. In the frame rotating with the resonator frequency, the Hamiltonian is modeled as (with $\hbar=1$)
\begin{align}
	H=\varepsilon(a^{\dagger2}+a^{2})-Ka^{\dagger2}a^{2},
	\label{eq:H_eff}
\end{align}
where $\varepsilon$ is the strength of the TPD and $K$ is the Kerr nonlinearity. Such a model can be implemented in various physical platforms. A particularly relevant realization is in circuit quantum electrodynamics (circuit-QED)~\cite{Krantz_2013,Lin2014,Grimm2020,Iyama2024,Xu2025}. The SPL is typically significant in such systems and must be accounted for. Additionally, we consider the ETPL, which has also been demonstrated in superconducting circuits~\cite{Leghtas2015,Touzard2018,PhysRevX.14.021019,Marquet2024}. We model the whole dynamics within the Markovian master equation framework:
\begin{align}
	\dot{\rho} = -i[H, \rho] + \kappa \mathcal{D}[a]\rho + \kappa_2 \mathcal{D}[a^2]\rho,
	\label{eq:master}
\end{align}
where $\mathcal{D}[O]\rho = O\rho O^\dagger - \frac{1}{2}(O^\dagger O \rho + \rho O^\dagger O)$ ($O=a,a^{2}$) is the Lindblad dissipator.
The term $\kappa \mathcal{D}[a]\rho$ describes the natural SPL, while $\kappa_2 \mathcal{D}[a^2]\rho$ represents the ETPL.

\section{\label{sec:level2}Dynamics of quantum metrological resources under SPL-induced decoherence}

To assess the metrological utility of the quantum states we can prepare, we quantify their performance in a parameter estimation protocol using the quantum Fisher information (QFI). For a perturbation generated by an operator $A$, the QFI associated with a state $\rho$ is defined as~\cite{Braunstein1994,Marti2024,Guo2024}
\begin{align}
	F_{Q}[\rho,A]=2\sum_{k,l}\frac{(\lambda_{k}-\lambda_{l})^{2}}{(\lambda_{k}+\lambda_{l})}|\langle k|A|l\rangle|^{2},
\end{align}
where $\lambda_k$ and $|k\rangle$ are the eigenvalues and eigenvectors of $\rho$, respectively, and the summation is taken over all $k, l$ such that $\lambda_k + \lambda_l > 0$.
When estimating the amplitude of a displacement, the generator of the perturbation is given by $A(\theta) = X \sin\theta + P \cos\theta$,
where $\theta$ specifies the direction of the displacement in phase space, and $X=(a^{\dagger}+a)/\sqrt{2}$ and $P=i(a^{\dagger}-a)/\sqrt{2}$ are the phase-space quadratures. We define the maximum QFI as
\begin{align}
	F_Q^{\text{max}} = \max_{\theta} F_Q[\rho, A(\theta)],
\end{align}
which quantifies the highest achievable sensitivity for a given state.
Notably, a coherent state yields $F_Q[|\alpha\rangle, A(\theta)] = 2$. Thus, if coherent states are considered as classical resources, any value $F_Q^{\text{max}} > 2$ indicates non-classical metrological advantage.
We compute $F_Q^{\text{max}}$ numerically by first obtaining the density matrix $\rho$ from the solution to the master equation~\eqref{eq:master}, and then maximizing $F_Q[\rho, A(\theta)]$ over all $\theta$.
To facilitate comparison across different parameter regimes, we introduce the quantum Fisher information gain (QFIG)
\begin{align}
	G_{Q}=10\log_{10}(F^{\text{max}}_{Q}/F_Q[|\alpha\rangle, A]),
\end{align}
which expresses the enhancement in decibels (dB) relative to the standard quantum limit set by coherent states. Any $G_{Q}>0$~dB signifies a nonclassical metrological advantage.

Our first goal is to gain deeper insight into the roles of Kerr nonlinearity and ETPL in sustaining quantum-enhanced metrological performance under SPL-induced decoherence. We systematically compare three dynamical scenarios.
All models are subject to the same SPL rate and driven by a TPD of strength $\varepsilon$ (i.e., $\varepsilon=1$ and $\kappa/\varepsilon=0.01$).
The key difference lies in the presence of Kerr nonlinearity and/or ETPL:
\begin{itemize}
	\item \textbf{Hybrid TPD-Kerr-ETPL model} ($K/\varepsilon = 0.25$, $\kappa_2/\varepsilon \geq 0$): The top panel of Fig.~\ref{Fig1}(a) shows $G_Q(t)$ for varying $\kappa_2/\varepsilon$ from 0 to 5, with fixed $K/\varepsilon = 0.25$. To illustrate the underlying state evolution, the middle and bottom panels display the Wigner function and photon number distribution $P_n$, respectively, evaluated at selected times ($\varepsilon t = 0.3,\,1.2,\,3,\,4.8,\,6$, and $90$) for the representative case $\kappa_2/\varepsilon = 0.5$. The diagonal elements of the density matrices obtained from Eq.~\eqref{eq:master} correspond to the Fock-state populations $P_n$, which represent the occupation probabilities of different photon number states~\cite{Marti2024}. We consider a protocol where the system initially prepared in the ground state of the harmonic oscillator, $|0\rangle$, evolves for $t \geq 0$ according to Eq.~\eqref{eq:master}. 
	
	\item \textbf{TPD-Kerr model} (with $\kappa_2/\varepsilon = 0$): The top panel of Fig.~\ref{Fig1}(b) shows $G_Q(t)$ for varying $K/\varepsilon = 0.05,\,0.1,\,0.25$, and $0.5$. The corresponding Wigner functions and $P_n$ distributions (middle and bottom panels) are shown for the representative case $K/\varepsilon = 0.25$ at the same set of time instants.
	
	\item \textbf{TPD-ETPL model} (with $K/\varepsilon = 0$): The top panel of Fig.~\ref{Fig1}(c) shows $G_Q(t)$ for varying $\kappa_2/\varepsilon = 0.1,\,0.2,\,0.5$, and 1. The corresponding Wigner functions and $P_n$ distributions (middle and bottom panels) are shown for the representative case $\kappa_2/\varepsilon = 0.5$ at the same set of time instants.
\end{itemize}

\subsection{Metrological dynamics of the TPD-Kerr-ETPL model under SPL-induced decoherence}

We first show that, in the hybrid TPD-Kerr-ETPL model, the presence of ETPL significantly mitigates the damped oscillations of the QFIG, $G_Q$, induced by SPL, thereby extending the useful metrological time window---defined as the non-oscillatory interval during which $G_Q$ remains above a practically relevant threshold. 
The ETPL profoundly mitigates---but does not eliminate---the detrimental impact of SPL. 
We will discuss these effects in more detail below.

\begin{figure*}
	\centering
	\includegraphics[width=\textwidth]{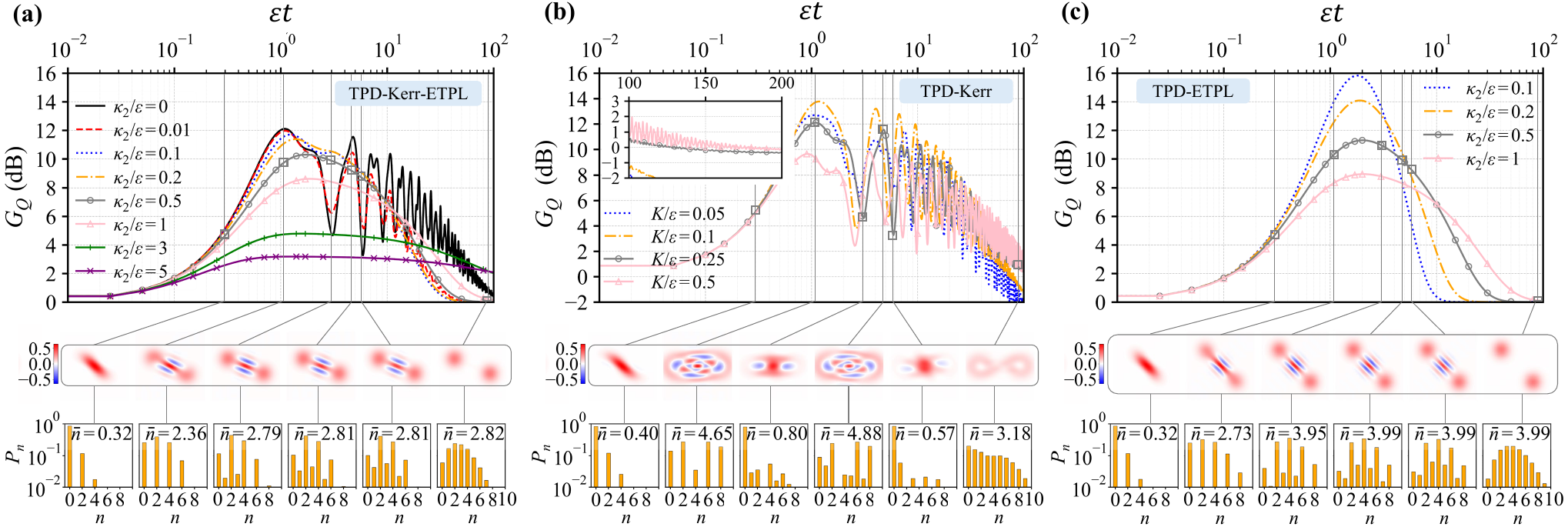}
	\caption{Comparison of metrological performance across three models: (a) the hybrid TPD-Kerr-ETPL model, (b) the TPD-Kerr model, and (c) the TPD-ETPL model. 
	(a)~The TPD-Kerr-ETPL model ($K/\varepsilon = 0.25$, $\kappa_2/\varepsilon \neq 0$): The top panel shows the time evolution of QFIG, $G_Q$, for varying $\kappa_2/\varepsilon$ from $0$ to $5$. 
	The middle and bottom panels display the Wigner function and photon-number distribution $P_n$, respectively, evaluated at selected times $\varepsilon t = 0.3,\,1.2,\,3,\,4.8,\,6$, and $90$ for $\kappa_2/\varepsilon = 0.5$.
	(b)~The TPD-Kerr model ($K/\varepsilon \neq 0$, $\kappa_2/\varepsilon = 0$): The top panel shows $G_Q(t)$ for $K/\varepsilon = 0.05$, $0.1$, $0.25$, and $0.5$. 
	The middle and bottom panels show the Wigner function and $P_n$ at the same time instants as in (a), for $K/\varepsilon = 0.25$.
	(c)~The TPD-ETPL model ($K/\varepsilon = 0$, $\kappa_2/\varepsilon \neq 0$): The top panel shows $G_Q(t)$ for $\kappa_2/\varepsilon = 0.1$, $0.2$, $0.5$, and $1$. 
	The middle and bottom panels depict the Wigner function and $P_n$ at the same time instants as in (a), for $\kappa_2/\varepsilon = 0.5$. In all panels, we set the SPL rate with $\kappa/\varepsilon = 0.01$.
	}\label{Fig1}
\end{figure*}

In the absence of ETPL (i.e., $\kappa_2 = 0$, indicated by the black solid line in the top panel of Fig.~\ref{Fig1}(a)), $G_Q$ first rises to $12$~dB at $\varepsilon t \approx 1.2$, then drops to a local minimum near $5$~dB before entering long-lived damped oscillations that persist for more than $\varepsilon t = 100$ (with $G_Q \approx 0.5$~dB at that time). 
The initial high-sensitivity time window---characterized by monotonic rise and decay without oscillation---ends when $G_Q$ first returns to this $\sim 5$~dB level. 
Accordingly, we take the interval over which $G_Q \geq 5$~dB and exhibits no oscillatory behavior (from $\varepsilon t \approx 0.3$ to $\varepsilon t \approx 3$) as a representative measure of the transient metrological window. 
This duration is too short for practical high-precision sensing applications.

This damped oscillation phenomenon can be explained as follows. 
In the absence of ETPL, the system evolves according to the master equation
\begin{align}
	\dot{\rho} = -i \left[ H_{\text{eff}}, \rho \right]  + \kappa a \rho a^{\dagger},
	\label{eq:H_a_eff}
\end{align}
where $H_{\text{eff}} = H - i \kappa a^{\dagger} a / 2$ is a non-Hermitian effective Hamiltonian. 
The approximately degenerate eigenstates of $H_{\text{eff}}$ are the coherent states $|\pm\alpha_0\rangle$~\cite{Puri2017}, with complex amplitude
\begin{align}
	\alpha_0 = r_0 e^{i\theta_0}, \quad 
	r_0 = \left( \frac{4\varepsilon^2 - \kappa^2/4}{4K^2} \right)^{1/4}, \quad
	\tan(2\theta_0) = \frac{\kappa}{\sqrt{16\varepsilon^2 - \kappa^2}},
\end{align}
valid in the regime $\kappa \ll 8K|\alpha_0|^2$.
The last term $\kappa a \rho a^{\dagger}$ in Eq.~\eqref{eq:H_a_eff} induces stochastic quantum jumps between even- and odd-parity cat states, i.e., $a |C^{+}_{\alpha_0}\rangle \propto |C^{-}_{\alpha_0}\rangle$.
This effect does not cause abrupt switching between distinct quantum phases but rather induces damped oscillations in $G_Q$.

In the long-time limit, the steady state is well approximated by a classical mixture of the two coherent states:
\begin{align}
	\rho_s \approx \frac{1}{2} \left( |\alpha_0\rangle\langle\alpha_0| + |{-}\alpha_0\rangle\langle{-}\alpha_0| \right).
	\label{eq:mix_state}
\end{align}
This state exhibits no quantum coherence between $|\alpha_0\rangle$ and $|{-}\alpha_0\rangle$, explaining why $G_Q$ asymptotically approaches a small residual value (less than $0.5$~dB by $\varepsilon t = 100$). 
Such a residual metrological gain is too weak and unstable for practical high-precision sensing.

Therefore, in the absence of ETPL---corresponding to the TPD-Kerr model---the damped oscillations of $G_Q$ and the narrow useful metrological time window (i.e., the transient nature of the metrological advantage) render the TPD-Kerr resonator unsuitable for practical high-precision sensing applications.

Introducing ETPL significantly mitigates these oscillations and substantially extends the useful metrological time window.
For weak ETPL (e.g., $\kappa_2/\varepsilon = 0.01$), the oscillation amplitudes are reduced.
For moderate ETPL (e.g., $\kappa_2/\varepsilon = 0.1$), the oscillations almost disappear.
However, because SPL continuously leaks population into the odd-parity subspace, quantum coherence cannot be maintained indefinitely. 
Consequently, $G_Q$ eventually decays monotonically to zero, reaching $G_Q = 0$ around $\varepsilon t \approx 60$---significantly earlier than in the $\kappa_2 = 0$ case, where $G_Q$ remains above $0.5$~dB even at $\varepsilon t = 100$.
Most importantly, the duration over which $G_Q \geq 5$~dB is extended by more than an order of magnitude (e.g., from $\varepsilon t \approx 0.3$ to $\varepsilon t \approx 20$ for $\kappa_2/\varepsilon = 1$), and the entire decay profile remains stable and easily trackable. 
We emphasize that the choice of $5$~dB is illustrative---it corresponds to the first return-to-baseline level in the $\kappa_2 = 0$ case (i.e., the TPD-Kerr model)---and our qualitative conclusion (dramatic window extension under ETPL) holds for any reasonable threshold below $8$~dB.

Counterintuitively, further increasing $\kappa_2/\varepsilon$ surpasses $0.1$ slows the long-time decay: for $\kappa_2/\varepsilon \geq 0.2$, $G_Q$ drops smoothly from its peak without oscillations but takes longer to reach zero than in the $\kappa_2/\varepsilon = 0.1$ case, thereby further extending the effective metrological time window.
This non-monotonic dependence arises because strong ETPL confines the system more tightly within the even-parity manifold, mitigating the effects of SPL-induced decoherence.

This picture is corroborated by the Wigner functions and photon-number distributions $P_n$ (middle and bottom panels of Fig.~\ref{Fig1}(a)), evaluated at $\varepsilon t = 0.3,\,1.2,\,3,\,4.8,\,6$, and $90$ for $\kappa_2/\varepsilon = 0.5$.
The evolution of $G_Q$ is accompanied by a transition among Gaussian, non-Gaussian, and mixed states:
(i) The system first develops Gaussian features; for example, a squeezed state exhibiting an elliptical Wigner function appears at $\varepsilon t \approx 0.3$, serving as a quantum metrological resource.
(ii) As the system evolves, non-Gaussian states emerge: an even-parity cat state forms at $\varepsilon t \approx 1.2$ and persists over an extended period, coinciding with the effective metrological time window and demonstrating a metrological advantage over Gaussian resources.
(iii) By $\varepsilon t = 90$, however, the Wigner function collapses into two disconnected positive lobes with no interference fringes, and $P_n$ becomes a near-Gaussian distribution centered at $n = 3$--$4$ with small but non-negligible odd-$n$ populations. These features confirm that SPL gradually erodes parity purity and quantum coherence, driving the system toward a classical mixture state (i.e., Eq.~\eqref{eq:mix_state}).

Thus, ETPL does not prevent SPL-induced decoherence, but rather mitigates SPL-induced decoherence so as to render the quantum-enhanced sensing capability more robust and predictable. 
By converting irregular damped oscillations into an easily trackable monotonic decay and extending the high-sensitivity window, ETPL provides a crucial practical advantage for real-world quantum metrology, where SPL is unavoidable.

\subsection{\label{sec:Comparison}Comparison of the TPD-Kerr and TPD-ETPL models under SPL-induced decoherence}

Having established that the hybrid TPD-Kerr-ETPL model effectively mitigates SPL-induced oscillations and extends the metrological time window through the stabilization of non-Gaussian even-parity cat states, we now turn to a comparative analysis of the two canonical mechanisms for generating non-Gaussian cat states: the Hamiltonian protection via TPD and Kerr nonlinearity (TPD-Kerr model), and the dissipative stabilization via TPD and ETPL (TPD-ETPL model). 
Specifically, we examine whether $G_Q$ is equivalent when both models are tuned with identical coherent-state amplitude $|\alpha|$ under the same SPL rate. 
The corresponding dynamics for the TPD-Kerr ($K \neq 0$, $\kappa_2 = 0$) and TPD-ETPL ($K = 0$, $\kappa_2 \neq 0$) models are shown in Figs.~\ref{Fig1}(b) and \ref{Fig1}(c), respectively.
The parameters in Figs.~\ref{Fig1}(b) and (c) are chosen such that $\kappa_2/K = 2$, ensuring that the effective cat size---characterized by the amplitude $|\alpha| = \sqrt{\varepsilon/K} = \sqrt{\varepsilon/(\kappa_2/2)}$---is identical in both the TPD-Kerr and TPD-ETPL models. This enables a fair assessment of whether ETPL provides a genuine advantage over passive nonlinear protection in preserving quantum-enhanced metrological sensitivity under the condition of SPL-induced decoherence.
Furthermore, by comparing Fig.~\ref{Fig1}(a) (TPD-Kerr-ETPL model) with Fig.~\ref{Fig1}(c) (TPD-ETPL model), we isolate the role of Kerr nonlinearity in shaping the metrological response and coherence dynamics in the presence of ETPL.

Both the TPD-Kerr and TPD-ETPL models achieve similar evolution trends within $\varepsilon t \lesssim 1.2$. 
However, their subsequent dynamics diverge significantly.
In the TPD-Kerr model (Fig.~\ref{Fig1}(b)), the SPL-induced stochastic parity jumps result in long-lived damped oscillations in $G_Q$. 
For moderate $K$, $G_Q$ eventually drops below $0$~dB (see the inset in Fig.~\ref{Fig1}(b))---indicating metrological performance worse than the standard quantum limit. 
The Wigner function is elongated along the real quadrature ($x$-axis), consistent with a real-valued coherent amplitude ($\alpha \in \mathbb{R}$).
The accompanying Wigner functions and photon-number distributions $P_n$ exhibit oscillatory dynamics: non-Gaussian features such as Wigner negativity and bimodal structure periodically emerge and decay, closely tracking the peaks and troughs of $G_Q$.
By $\varepsilon t = 90$, the system approaches a classical mixture state, with $P_n$ becoming nearly uniform across even and odd photon numbers.

In contrast, the TPD-ETPL model (Fig.~\ref{Fig1}(c)) leverages engineered loss to actively stabilize the even-parity manifold. 
$G_Q$ decays smoothly without oscillations. 
Although $G_Q$ ultimately vanishes due to residual SPL-induced decoherence, the effective high-sensitivity time window (e.g., $G_Q \geq 5$~dB) is significantly extended---for instance, for $\kappa_2/\varepsilon = 0.5$, $G_Q \geq 5$~dB persists from $\varepsilon t \approx 0.3$ to $\varepsilon t \approx 14$, far longer than the interval from $\varepsilon t \approx 0.3$ to $\varepsilon t \approx 3$ observed in the TPD-Kerr model with $K/\varepsilon = 0.25$.
The Wigner function is oriented along the diagonal ($-\pi/4$ direction), reflecting the complex steady-state amplitude $\alpha$ with phase $e^{-i\pi/4}$.

Furthermore, by comparing the TPD-ETPL model in Fig.~\ref{Fig1}(c) with the TPD-Kerr-ETPL model depicted in Fig.~\ref{Fig1}(a), the interplay between Kerr nonlinearity $K$ and ETPL rate $\kappa_2$ becomes evident. 
For fixed $K/\varepsilon = 0.25$ in Fig.~\ref{Fig1}(a), at small $\kappa_2/\varepsilon$ (e.g., $\kappa_2/\varepsilon = 0.1$), the hybrid TPD-Kerr-ETPL model exhibits a lower peak and an earlier decay to zero, reflecting the detrimental effect of Kerr-induced dynamics.
As $\kappa_2/\varepsilon$ increases from $0.1$ to $1$, this discrepancy gradually diminishes: for $\kappa_2/\varepsilon = 1$, both the TPD-ETPL and hybrid TPD-Kerr-ETPL models reach $G_Q = 0$ at nearly the same time ($\varepsilon t \approx 90$), despite the latter having a slightly reduced peak.
This convergence demonstrates that strong ETPL dominates over parasitic Kerr nonlinearity, effectively restoring the smooth, easily trackable decay characteristic of the TPD-ETPL model.
Thus, while Kerr nonlinearity degrades the ETPL performance at weak ETPL strengths, it becomes negligible when $\kappa_2 \gtrsim K$. This establishes a practical design principle: to harness the stability of ETPL in real devices, one need not eliminate $K$ entirely---only ensure that $\kappa_2$ is sufficiently large compared to $K$.

Thus, our comparative analysis demonstrates that ETPL offers a decisive advantage over the TPD-Kerr model in preserving metrological stability under SPL. While the TPD-Kerr model suffers from persistent damped oscillations and a narrow high-sensitivity window, the TPD-ETPL model achieves smooth, easily trackable decay of $G_Q$ and extends the useful metrological time window by more than an order of magnitude.
Crucially, when both models are matched for equivalent cat-state amplitude, only ETPL ensures stable, easily trackable metrological performance under SPL.
Moreover, in the hybrid TPD-Kerr-ETPL model, strong ETPL ($\kappa_2 \gtrsim K$) effectively suppresses the detrimental effects of Kerr dynamics, restoring metrological performance close to that of the TPD-ETPL model.
This establishes a practical design principle for real-world quantum metrology: rather than eliminating intrinsic Kerr nonlinearities, one can harness ETPL to dominate the dynamics and ensure robust, long-lived quantum-enhanced sensing.

\section{\label{sec:Squeezing}Interplay between quantum squeezing and metrological sensitivity under SPL-induced decoherence}

In this section, we address two distinct questions: the role of Kerr nonlinearity, SPL and ETPL in quantum squeezing processes, and the potential interplay between quantum sensing and squeezing.
We examine these questions within the same three models as before: the hybrid TPD-Kerr-ETPL model (Fig.~\ref{Fig2}(a)), the TPD-Kerr model (Fig.~\ref{Fig2}(b)), and the TPD-ETPL model (Fig.~\ref{Fig2}(c)).
All models are subject to the same SPL rate and driven by an identical TPD of strength $\varepsilon$ (i.e., $\varepsilon=1$ and $\kappa/\varepsilon=0.01$).

To quantify the squeezing level, we adopt the phase-space quadratures as 
$X=(a^{\dagger}+a)/\sqrt{2}$ and $P=i(a^{\dagger}-a)/\sqrt{2}$,
and the variance along an arbitrary direction $\theta$ as~\cite{Marti2024,Guo2024} 
\begin{align}
	V(\theta) = \mathrm{Var}[X\cos\theta + P\sin\theta].
\end{align}
For the harmonic oscillator ground state (vacuum), the variance is $V_{\text{GS}} = 1/2$, independent of $\theta$. 
A variance $V(\theta) < V_{\text{GS}}$ therefore indicates quantum squeezing.
We define the minimum variances as 
\begin{align}
	V_{\text{min}} = \min_{\theta} V(\theta),
\end{align} 
which corresponds to the squeezed quadrature. The corresponding squeezing level in decibels (dB) is quantified as 
\begin{align}
	S = -10\,\log_{10}(V_{\text{min}}/V_{\text{GS}}).
\end{align}
Here, $S > 0$ indicates squeezing (noise below the vacuum level), $S = 0$ corresponds to the vacuum state (no squeezing), and $S < 0$ indicates that no squeezing is present and the minimum noise exceeds the vacuum level.

\subsection{Squeezing dynamics of the TPD-Kerr-ETPL model under SPL-induced decoherence}

\begin{figure*}
	\centering
	\includegraphics[width=\textwidth]{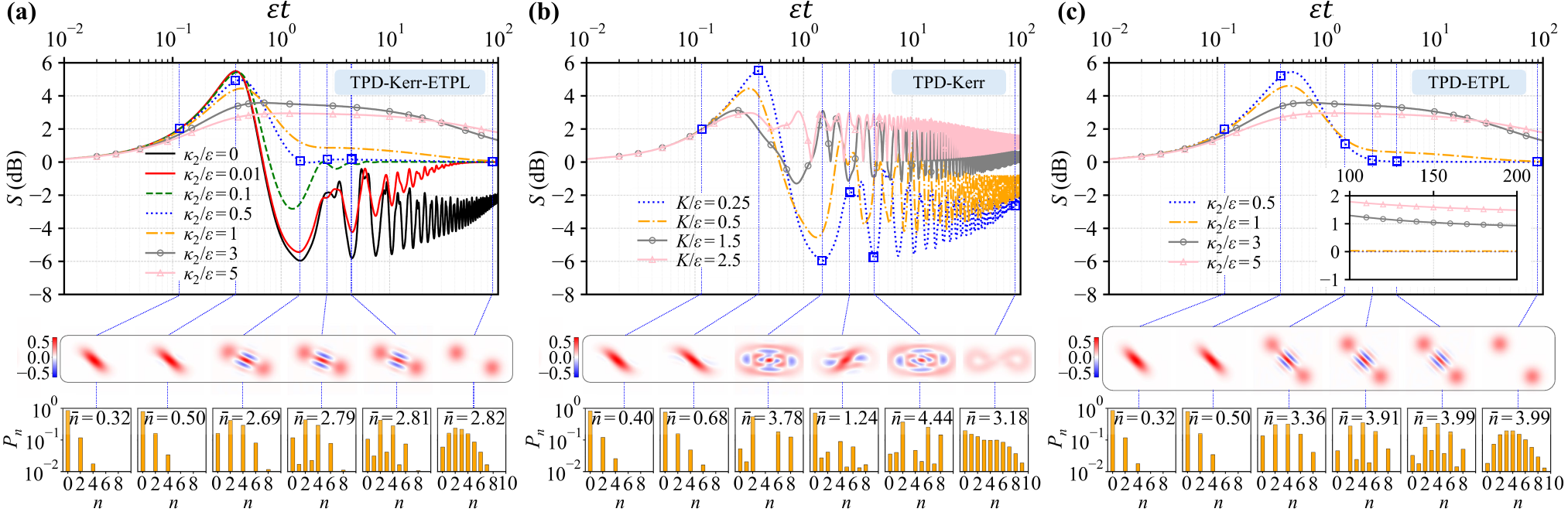}
	\caption{Comparison of squeezing performance across three models: (a) the hybrid TPD-Kerr-ETPL, (b) the TPD-Kerr model, and (c) the TPD-ETPL model. 
	(a)~The TPD-Kerr-ETPL model ($K/\varepsilon = 0.25$, $\kappa_2/\varepsilon \neq 0$): The top panel shows the time evolution of quantum squeezing level $S$ for $\kappa_2/\varepsilon$ varying from $0$ to $5$. 
	The middle and bottom panels display the Wigner function and photon-number distribution $P_n$, respectively, evaluated at selected times $\varepsilon t = 0.3,\,0.38,\,1.6,\,2.7,\,4.5$, and $90$ for $\kappa_2/\varepsilon = 0.5$.
	(b)~The TPD-Kerr model ($K/\varepsilon \neq 0$, $\kappa_2/\varepsilon = 0$): The top panel shows the time evolution of $S$ for $K/\varepsilon = 0.25$, $0.5$, $1.5$, and $2.5$. 
	The middle and bottom panels show the Wigner function and $P_n$ at the same time instants as in (a), for $K/\varepsilon = 0.25$.
	(c)~The TPD-ETPL model ($K/\varepsilon = 0$, $\kappa_2/\varepsilon \neq 0$): The top panel shows the time evolution of $S$ for $\kappa_2/\varepsilon = 0.5$, $1$, $3$, and $5$. 
	The middle and bottom panels depict the Wigner function and $P_n$ at the same time instants as in (a), for $\kappa_2/\varepsilon = 0.5$. In all panels, we set $\kappa/\varepsilon=0.01$.
	}\label{Fig2}
\end{figure*}

We first show that, in the hybrid TPD-Kerr-ETPL model, ETPL plays a multifaceted role in shaping quantum metrological resources: (i) it suppresses the long-lived antisqueezing ($S<0$) oscillations induced by SPL, ensuring rapid convergence to the vacuum level ($S=0$); (ii) it significantly extends the effective squeezing window---defined as the duration over which $S \geq 3$~dB (a representative threshold commonly used to signify practically useful squeezing)---by up to an order of magnitude; and (iii) it engineers a temporal separation between Gaussian squeezing (short-time, high $S$) and non-Gaussian cat-state formation (intermediate-time, high $G_Q$), thereby enabling both initial noise suppression and sustained quantum-enhanced sensing. This synergistic control over the resource landscape complements the stabilization of $G_Q$ discussed earlier and underscores ETPL as a powerful tool for engineering robust metrological protocols.

In the absence of ETPL ($\kappa_2 = 0$), as depicted in Fig.~\ref{Fig2}(a), $S$ rises rapidly to a peak of $S\approx 5.5$~dB at $\varepsilon t \approx 0.38$, then plunges into strong antisqueezing ($S \approx -6$~dB) and undergoes oscillations that remain predominantly negative, approaching $S \approx -2$~dB at $\varepsilon t = 100$ while slowly trending toward zero. 
Such prolonged antisqueezing implies that the system remains in a high-noise state for most of its evolution, degrading metrological utility.

Introducing weak ETPL ($\kappa_2/\varepsilon = 0.01$ or $0.1$) preserves the initial squeezing peak but suppresses the depth of antisqueezing.
The key advantage lies in significantly mitigating the persistent oscillations that would otherwise keep the system in a high-noise ($S < 0$) state indefinitely. 
Instead, the system reliably relaxes to $S = 0$ within a finite time---achieved by $\varepsilon t \approx 60$ and $\varepsilon t \approx 10$, respectively.

For stronger ETPL ($\kappa_2/\varepsilon \geq 0.5$), the dynamics change qualitatively: the peak squeezing occurs later and with reduced magnitude, but crucially, $S$ remains non-negative throughout its decay and smoothly approaches zero without oscillations. 
At even higher loss strengths ($\kappa_2/\varepsilon = 3$ or $5$), the peak squeezing is further reduced and occurs later, yet $S$ retains a significant positive value even at $\varepsilon t = 100$. 
Consequently, the effective squeezing window---defined here as the interval where $S \geq 3$~dB---is dramatically extended: for $\kappa_2/\varepsilon = 3$, it spans from $\varepsilon t \approx 0.28$ to $\varepsilon t \approx 13$, whereas in the $\kappa_2 = 0$ case it lasts only from $\varepsilon t \approx 0.28$ to $\varepsilon t \approx 0.56$.
Thus, ETPL not only eliminates detrimental long-time antisqueezing but also enables tunable, sustained squeezing---both critical for practical quantum metrology.

Notably, the temporal relationship between squeezing and sensing reveals a rich picture.
For weak or absent ETPL, the peak of $S$ occurs around $\varepsilon t \approx 0.4$, coinciding with the initial rise of $G_Q$ in Fig.~\ref{Fig1}(a); however, $G_Q$ continues to increase and peaks near $\varepsilon t \approx 1.2$, precisely when $S$ has already become negative (no squeezing). 
In contrast, for $\kappa_2/\varepsilon \gtrsim 0.5$, the squeezing peak shifts to $\varepsilon t \approx 1$, aligning more closely with the onset of the high-$G_Q$ regime, and the squeezing window broadens significantly.

To elucidate the underlying state evolution, we examine the representative case $\kappa_2/\varepsilon = 0.5$: at $\varepsilon t = 0.3$ and $0.38$, the Wigner function exhibits a Gaussian elliptical shape characteristic of a squeezed vacuum, with the latter time corresponding to the peak $S$.
By $\varepsilon t \approx 1.2$, when $G_Q$ reaches its maximum, the state has evolved into a non-Gaussian even-parity cat state, as evidenced by clear interference fringes in the Wigner function, even though $S$ has already decayed to near zero. 
At long times ($\varepsilon t = 90$), both $G_Q$ and $S$ vanish as the system relaxes to a classical mixture.

This progression demonstrates that short-time metrological enhancement is driven by Gaussian squeezing ($S > 0$, $G_Q$ rising), whereas long-time advantage stems from non-Gaussian even-parity cat states ($S \approx 0$, $G_Q$ sustained)---highlighting the superior sensing capability of engineered non-Gaussian resources over pure Gaussian squeezing. Moreover, ETPL not only eliminates detrimental antisqueezing but also actively tailors the temporal profile of quantum resources: it broadens the useful squeezing window, aligns the peak of $G_Q$ with a stable post-squeezing regime, and ensures smooth decay to a vacuum noise level. Together, these features establish engineered loss as a versatile strategy for optimizing both the quality and duration of quantum-enhanced metrology in realistic noisy environments.

\subsection{Comparison of squeezing dynamics in the TPD-Kerr and TPD-ETPL models under SPL-induced decoherence}

Having established that the hybrid TPD-Kerr-ETPL model not only suppresses SPL-induced antisqueezing oscillations but also extends the effective squeezing window through dissipative stabilization of even-parity cat states, we now turn to a comparative analysis of the two canonical mechanisms for generating quantum squeezing and non-Gaussian resources: the Hamiltonian protection via TPD and Kerr nonlinearity (the TPD-Kerr model), and the dissipative stabilization via TPD and ETPL (the TPD-ETPL model).  
Specifically, we examine whether the squeezing level $S$ exhibits identical dynamics when both models are tuned to produce cat states of identical coherent-state amplitude $|\alpha|$ under same SPL rate $\kappa$.  
The corresponding dynamics for the TPD-Kerr ($K \neq 0$, $\kappa_2 = 0$) and TPD-ETPL ($K = 0$, $\kappa_2 \neq 0$) models are shown in Figs.~\ref{Fig2}(b) and \ref{Fig2}(c), respectively.  
The parameters in Figs.~\ref{Fig2}(b) and (c) are chosen such that $\kappa_2 / K = 2$, ensuring that the effective cat size is identical in both models. This enables a fair assessment of whether ETPL provides a genuine advantage over passive Kerr-based protection in preserving useful quantum squeezing under realistic noise conditions.  
Furthermore, by comparing Fig.~\ref{Fig2}(a) (hybrid TPD-Kerr-ETPL model) with Fig.~\ref{Fig2}(c) (TPD-ETPL model), we isolate the role of Kerr nonlinearity in shaping the temporal profile and stability of squeezing in the presence of engineered dissipation.

In the TPD-Kerr model where $\kappa_2=0$ (Fig.~\ref{Fig2}(b)), quantum squeezing arises from Hamiltonian nonlinear dynamics, but it is highly vulnerable to SPL, resulting in irregular oscillations and limited operational windows. This behavior corresponds to the $\kappa_2/\varepsilon=0$ case in Fig.~\ref{Fig2}(a).
Despite achieving comparable initial squeezing, the absence of ETPL renders the resource transient and unreliable for sustained sensing.
Specifically, for $\kappa_2 = 0$ and varying $K/\varepsilon = 0.25,\,0.5,\,1.5,\,2.5$, as shown in Fig.~\ref{Fig2}(b), all cases exhibit an initial squeezing peak followed by damped oscillations. 
However, the long-time behavior diverges: for small $K/\varepsilon$ ($0.25$ and $0.5$), $S$ remains predominantly negative; for intermediate $K/\varepsilon$ ($1.5$), it oscillates around zero; and for large $K/\varepsilon$ ($2.5$), it settles into a weakly squeezed regime ($S \approx 1$~dB).

Despite these differences, the effective squeezing window remains narrow in all cases, and $S$---like $G_Q$ in Fig.~\ref{Fig1}(b)---undergoes irregular oscillations that preclude stable metrological operation. 
This reinforces the conclusion drawn from $G_Q$: the TPD-Kerr scheme suffers from transient, unstable quantum resources, limiting its metrological utility in realistic noisy environments.

In contrast, the squeezing dynamics of the TPD-ETPL model with $K=0$ in the top panel of Fig.~\ref{Fig2}(c) demonstrate that engineered loss alone is sufficient to generate and stabilize squeezing with dynamics nearly identical to the hybrid TPD-Kerr-ETPL case. 
The middle and bottom panels in Fig.~\ref{Fig2}(c) display squeezing trajectories for $\kappa_2/\varepsilon = 0.5,\,1,\,3,\,5$ that closely mirror those in Fig.~\ref{Fig2}(a) at the same $\kappa_2/\varepsilon$.
This confirms that when $\kappa_2 \gtrsim K$, the Kerr nonlinearity becomes negligible, and ETPL dominates the metrological response---providing a clean platform for noise-resilient quantum resource engineering.
Moreover, even when the TPD-Kerr and TPD-ETPL models in Figs.~\ref{Fig2}(b--c) are tuned to the same amplitude $|\alpha|$, their dynamical responses to identical SPL differ markedly.

Therefore, our comparative analysis of squeezing dynamics reveals that ETPL provides a robust and superior mechanism for stabilizing quantum resources compared to the TPD-Kerr model.  While the TPD-Kerr model generates transient squeezing that is highly susceptible to SPL-induced oscillations and rapid degradation, the TPD-ETPL model---despite lacking any Hamiltonian nonlinearity---achieves smooth, non-oscillatory squeezing decay and significantly extends the useful squeezing time window.
Crucially, when both models are matched for equivalent cat-state amplitude, only ETPL ensures stable, easily trackable metrological performance under SPL. Moreover, the near-identical squeezing trajectories between the TPD-ETPL and hybrid TPD-Kerr-ETPL models confirm that the strong ETPL effectively mitigates the detrimental function of Kerr nonlinearities.
These findings establish ETPL not merely as a complementary tool, but as a dominant strategy for engineering broad squeezing time windows in practical quantum metrology platforms.

\section{\label{sec:level4}Discussion and conclusion}

We now assess the experimental feasibility of our proposal supported by recent advances in superconducting circuit QED. State-of-the-art platforms now achieve ETPL rates $\kappa_2$ up to several MHz~\cite{Touzard2018,PhysRevX.14.021019,Marquet2024}, exceeding the typical SPL rates ($\kappa \sim 10$--$100$~kHz) by over two orders of magnitude. This wide separation enables a dimensionless regime explored in our study, where $\kappa_2/\varepsilon \in [0,5]$ and $\kappa/\varepsilon = 10^{-2}$.
Notably, our choice of a relatively strong SPL rate ($\kappa/\varepsilon = 10^{-2}$) reflects a more realistic noise environment compared to recent Kerr-cat qubit experiments, which typically operate in the range $\kappa/\varepsilon \sim 10^{-4}$--$10^{-3}$ with comparable Kerr-to-drive ratios ($K/\varepsilon \approx 0.15$--$1.01$)~\cite{Grimm2020,Iyama2024,Xu2025}. Our analysis thus addresses a more challenging---and arguably more practical---regime where noise cannot be neglected, and demonstrates that ETPL remains effective even under such conditions.

Furthermore, the loss engineering approach discussed here can be generalized to other bosonic systems where controlled two-phonon loss can be engineered. 
For example, the generation of mechanical Schr\"odinger cat states~\cite{doi:10.1126/science.adf7553} and the realization of strong squeezing below the zero-point fluctuations in a gigahertz-frequency phonon mode of a high-overtone bulk acoustic-wave resonator with tunable nonlinearity have been achieved~\cite{Marti2024,PhysRevLett.134.180801}. 
In such mechanical systems, two-phonon loss can similarly stabilize nonclassical states and sustain enhanced quantum sensing performance.
Two-phonon loss has been experimentally realized in several mechanical systems. For example, the mechanical van der Pol (vdP) oscillator with two-phonon loss can be implemented in two prominent experimental platforms: 
(i) trapped-ion systems, where the two-phonon loss can be realized via laser excitation to its red motional sideband by removing two phonons at a time~\cite{PhysRevLett.111.234101,PhysRevLett.112.094102,PhysRevLett.117.073601,PhysRevLett.119.133601,PhysRevA.95.041802,PhysRevLett.120.163601,Li2025}, and 
(ii) optomechanical ``membrane-in-the-middle''
systems~\cite{PhysRevA.46.2668,Jayich_2008,Thompson2008,10.1063/5.0026286} or the mechanical self-oscillation in cavity optomechanical systems~\cite{PhysRevLett.96.103901,PhysRevLett.95.033901,PhysRevLett.101.133903}, in which the nonlinear two-phonon loss can be implemented by exploiting a laser red-detuned by twice the mechanical frequency to couple to the two-phonon sideband.
These platforms offer promising avenues for extending the present results to quantum-enhanced mechanical sensing and metrology.

To summarize, we have investigated a two-photon-driven Kerr nonlinear oscillator subject to unavoidable SPL---a common noise channel in quantum hardware---with additional ETPL introduced to enhance metrological performance. This setup allows us to investigate how metrological performance depends on the interplay between Kerr nonlinearity and engineered loss. Specifically, we compare three variants of the same physical setup: (i) a TPD-Kerr model, where quantum resources are generated by TPD and Kerr nonlinearity; (ii) a TPD-ETPL model, which relies on the TPD and ETPL to stabilize states; and (iii) a hybrid TPD-Kerr-ETPL model, combining both mechanisms under identical driving conditions.
We find that in the TPD-Kerr case, SPL induces long-lived oscillations in sensing performance, restricting useful operation to extremely short time windows.
In contrast, even weak ETPL suppresses these oscillations completely, yielding a smooth decay and extending the high-sensitivity window by over an order of magnitude.
When the engineered loss dominates intrinsic nonlinearity, the hybrid system converges quantitatively to the TPD-ETPL limit.
Moreover, even when the TPD-Kerr and TPD-ETPL models are tuned to the same amplitude, their dynamical responses to identical SPL differ markedly.
Furthermore, we have uncovered a temporal hierarchy of quantum resources: early enhancement of metrological sensitivity stems from Gaussian squeezing, while sustained advantage relies on dissipatively stabilized non-Gaussian even-parity cat states. Critically, this stabilization occurs fully autonomously---without encoding, feedback, or real-time control. Our results thus establish a general design principle: tailored loss, combined with coherent driving, can actively preserve metrologically useful quantum resources against dominant noise, paving the way toward robust, scalable quantum sensors in superconducting circuits, optomechanics, and trapped-ion platforms.

\section{\label{Methods}Methods}

\subsection{\label{Method-A}Generation of ETPL}
In this section, we outline the implementation of ETPL.
ETPL has been experimentally realized in several superconducting circuit architectures~\cite{Mirrahimi_2014,Leghtas2015,Touzard2018,Lescanne2020,Berdou2023,Reglade2024,Marquet2024}. 
A common approach employs an auxiliary mode, referred to as the buffer, to mediate the interaction between the memory mode and the environment~\cite{Leghtas2015,Touzard2018,Lescanne2020,Berdou2023,Reglade2024,PhysRevX.14.021019,Marquet2024}. 
The buffer mode can exhibit a high loss rate, with $\kappa_b/2\pi \approx 40$~MHz~\cite{Marquet2024}.
The memory ($a$) and buffer ($b$) modes are coupled via a nonlinear element implementing a two-photon exchange Hamiltonian $H_{ab} = g_2 (a^2 b^\dagger + a^{\dagger 2} b)$, achieved through three-wave mixing under the resonance condition $2\omega_a = \omega_b$~\cite{PhysRevX.14.021019,Marquet2024}.

In the regime where $\kappa_b \gg 8g_2 |\alpha|$, photons in the buffer mode decay into the environment much faster than they are converted into pairs in the memory mode. 
This separation of timescales enables the adiabatic elimination of the buffer mode, resulting in an effective ETPL described by the jump operator $\sqrt{\kappa_2}\, a^2$, with the effective rate given by $\kappa_2 = 4g_2^2 / \kappa_b$.
Crucially, the coupling strength $g_2$---and thus $\kappa_2$---can be tuned experimentally. In the autoparametric cat qubit architecture~\cite{PhysRevX.14.021019,Marquet2024}, $g_2$ depends mainly on the magnetic flux detuning from the sweet spot, the zero-point phase fluctuations of the modes, the Josephson junction energy, and the precise frequency matching ($\omega_b = 2\omega_a$ is required to maintain large $g_2$, as small detunings reduce the effective coupling).

Recent experiments have demonstrated $\kappa_2/2\pi$ up to $2.16$~MHz---over 150 times larger than typical SPL rates---by operating at the optimal flux point $\phi_\text{on}$~\cite{Marquet2024}. At another flux point $\phi_\text{off}$, ETPL is effectively switched off. Other platforms have achieved $\kappa_2/2\pi \sim 200$--$280$~kHz with $\kappa_2/\kappa \gtrsim 10^2$~\cite{Touzard2018,PRXQuantum.3.010329,PhysRevX.9.041053}, confirming the feasibility of strong ETPL for quantum metrology.

\subsection{\label{Method-B}Long-time steady-state solutions of the TPD-Kerr-ETPL model without SPL}
The long-time steady states of a two-photon-driven Kerr resonator with ETPL are governed by the master equation
\begin{align}
	\dot{\rho} = -i[H, \rho] + \kappa_2 \mathcal{D}[a^2]\rho,
	\label{eq:masterETPL}
\end{align}
where $H = \varepsilon (a^{\dagger 2} + a^2) - K a^{\dagger 2} a^2$.
Steady-state solutions satisfy $\dot{\rho} = 0$, leading to the operator equation:
\begin{align}
	-i\varepsilon a^{2}\rho+i\rho a^{2}\varepsilon-i\varepsilon a^{\dag2}\rho+i\rho a^{\dag2}\varepsilon+iKa^{\dag2}a^{2}\rho-iK\rho a^{\dag2}a^{2}-\frac{1}{2}\kappa_{2}a^{\dag2}a^{2}\rho+\kappa_{2}a^{2}\rho a^{\dag2}-\frac{1}{2}\kappa_{2}\rho a^{\dag2}a^{2}=0.
	\label{eq:steadystate}
\end{align}
To solve this equation, we follow an approach analogous to that used in the absence of Kerr nonlinearity~\cite{Gerry1993}, adapted to include the $K a^{\dagger 2} a^2$ term. 
Assuming a pure steady state $\rho(\infty) = |z\rangle\langle z|$, where $|z\rangle$ is an eigenstate of the annihilation operator squared, $a^2 |z\rangle = z |z\rangle$, such that $a^{2}\rho(\infty)=z\rho(\infty)$  and $\rho(\infty) a^{\dag2}=z^{*}\rho(\infty) $.
Substituting into Eq.~(\ref{eq:steadystate}) and evaluating the commutators, one obtains the condition:
\begin{align}
	z = \varepsilon/(K + i \kappa_2 / 2).
	\label{eq:z_solution}
\end{align}
Thus, the amplitude of the coherent components is $\alpha = \sqrt{z} = \sqrt{\varepsilon / (K + i \kappa_2 / 2)}$, and the steady states are superpositions within the manifold spanned by $|\pm\alpha\rangle$.
Due to the even-parity structure of all terms in Eq.~(\ref{eq:masterETPL})---involving only even powers of $a$ and $a^\dagger$---the excitation parity is conserved. 
Consequently, the system evolves into either the even-parity cat state $|\mathcal{C}_\alpha^+\rangle \propto |\alpha\rangle + |{-}\alpha\rangle$ or the odd-parity state $|\mathcal{C}_\alpha^-\rangle \propto |\alpha\rangle - |{-}\alpha\rangle$, depending on the initial state's parity.
For example, an initial vacuum state $|0\rangle\langle 0|$ evolves into $|\mathcal{C}_\alpha^+\rangle$, while $|1\rangle\langle 1|$ relaxes to $|\mathcal{C}_\alpha^-\rangle$.
At early times, when ETPL is negligible, the dynamics are governed by the Hamiltonian $H$, under which the vacuum state $|0\rangle$ evolves into a squeezed vacuum and $|1\rangle$ into a squeezed number state. 
In the long-time limit, the balance between TPD and loss drives the system toward the cat-state manifold, demonstrating the role of ETPL in stabilizing non-Gaussian steady states.

To elucidate the stabilization mechanism, we rewrite the master equation in the non-Hermitian effective Hamiltonian formalism:
\begin{align}
	\dot{\rho} = -i \left( H_{\text{eff}} \rho - \rho H_{\text{eff}}^\dagger \right) + \kappa_2 a^2 \rho a^{\dagger 2},
	\label{eq:nonherm}
\end{align}
where $H_{\text{eff}} = H - i \kappa_2 a^{\dagger 2} a^2 / 2$. 
The imaginary Kerr-like term $-i \kappa_2 a^{\dagger 2} a^2 / 2$ introduces a photon-number-dependent damping that selects the amplitude $\alpha$, while the jump operator $a^2$ preserves the parity of the state, as $a^2 |\pm\alpha\rangle = \alpha^2 |\pm\alpha\rangle$ and $a^2 |\mathcal{C}_\alpha^\pm\rangle \propto |\mathcal{C}_\alpha^\pm\rangle$. 
Therefore, ETPL does not induce transitions between the even- and odd-parity cat states and is compatible with autonomous stabilization of encoded quantum information.

\Acknowledgements{We thank Yanyan Cai at Southern University of Science and Technolog for valuable discussions. This work was supported by the National Natural Science Foundation of China (Grant Nos. 12475015, 12274080, 12474356 and 11875108) and the Quantum Science and Technology-National Science and Technology Major Project (Grant No. 2021ZD0300200).}



\bibliographystyle{scis}
\bibliography{Ref}


\end{document}